\documentclass{aa}  

\usepackage{graphicx}
\usepackage{natbib}
\usepackage{mathtools}
\usepackage{multirow}
\usepackage{amsmath}
\usepackage{url}
\usepackage{txfonts}

\RequirePackage{color}
\usepackage{xcolor}
\definecolor{darkgreen}{rgb}{0.0, 0.4, 0.0}

\graphicspath{{./}{figures/}}

\begin{document}

   \title{Transverse waves observed in a fibril with the MiHI prototype}
    \titlerunning{A case study of transverse waves observed by MiHI}

     \author{E. Petrova \inst{1}
        \and T. Van Doorsselaere\inst{1}
        \and M. van Noort\inst{2}
        \and D. Berghmans\inst{3}
        \and J. S. Castellanos Durán\inst{2}
          }

   \institute{Centre for mathematical Plasma Astrophysics, Mathematics Department, KU Leuven, Celestijnenlaan 200B bus 2400, B-3001 Leuven, Belgium\\
        \email{elena.petrova@kuleuven.be}
        \and Max-Planck-Institut für Sonnensystemforschung, Justus-von-Liebig-Weg 3, D-37077 Göttingen, Germany
        \and Solar-Terrestrial Centre of Excellence – SIDC, Royal Observatory of Belgium, Ringlaan -3- Av. Circulaire, 1180 Brussels, Belgium}

   \date{Received; accepted}

  \abstract
   {Fine-scale structures of the solar chromosphere, particularly fibrils, are known to host various types of magnetohydrodynamic (MHD) waves that can transport energy to the corona. In particular, absorption features observed in the H$\alpha$ channel have been widely detected that exhibit transverse oscillations.}
   {We aimed to detect a high-frequency transverse oscillation in fibrils. }
   {We conducted a case study on a high-frequency transverse oscillation in a chromospheric fibril. A chromospheric fibril was observed on 24 August 2018, in the H$\alpha$ spectral line, with the prototype Microlensed Hyperspectral Imager (MiHI) at the Swedish 1-meter Solar Telescope. The MiHI instrument is an integral field spectrograph capable of achieving ultra-high resolution simultaneously in the spatial, temporal, and spectral domains.}
   {The detected oscillation characteristics include a period of 15 s and a displacement amplitude of 42 km. Using the bisector method, we derived Doppler velocities and determined that the polarisation of the oscillation was elliptical.}
   {The energy contained in the oscillation ranges from 390 to 2300 W/m$^2$, which is not sufficient to balance radiative losses of the chromosphere. }

   \maketitle

\section{Introduction} \label{sec:intro}

The chromosphere is a region in the solar atmosphere that contains various kinds of fine-scale structures such as spicules at the limb, relatively stable fibrils in active regions (ARs), along with shorter and more dynamic ones, and mottles in quiet Sun regions. The relationship between these structures, as well as their driving mechanisms, is not yet fully understood. A common phenomenon uniting these structures is the presence of various magnetohydrodynamic oscillations, particularly transverse modes. Those oscillations have been detected quite extensively in spicules \citep[e.g.][]{DePontieu2007,Bate2022}, mottles \citep[e.g.][]{Kuridze2012}, and rapid blue-shifted excursions \citep[RBEs, see for example][]{Sekse2013}. 

There is also an abundance of studies on oscillations in fibrils: sunspot super-penumbral fibrils \citep{Morton2021} with mean period of 754 s and displacement of 74 km; internetwork fibrils \citep{Mooroogen2017} with a mean period of 128 s and a displacement of 85 km; slender \ion{Ca}{II} H fibrils \citep{Jafarzadeh2017} with a period of 16-199 s and a displacement of 1-91 km; and quiet Sun fibrils \citep{Kwak2023} with a 7.5$\pm$5.6 min period for all observed waves.
In most cases, the oscillations in fibrils are referred to as Alfvénic waves, a term originally introduced to indicate that the observed wave modes share characteristics with the pure Alfvén mode. This category of waves also includes kink waves, which, according to \cite{goossens2012}, are of an Alfvénic character, since kink waves are almost incompressible and have a magnetic tension acting as a restoring force. However, this term is also used in a much wider range in the cases where the observed waves characteristics satisfy only some of the criteria for a pure Alfvén wave, as the full nature of the wave cannot be determined with certainty \cite{McIntosh2008, Chelpanov2024, Morton2025}.

Although the above-mentioned works focus primarily on statistical studies, there are also case studies that are devoted to the 3D motion of spicules, as researched by \cite{Sharma2017,Sharma2018,Shetye2021, Kianfar2022}. The study of the 3D motion is typically enabled by augmenting imaging observations with Doppler velocities, either observed with spectrographs or derived.  This sheds light on the complex motion of the features and aids in differentiating between two possible wave modes explaining the motion - torsional Alfvén waves and kink waves - by deducing the line-of-sight (LoS) component of the perturbed magnetic field. Moreover, if the detected waves are interpreted as kink waves, this allows one to determine one of their key characteristics, which is polarisation. 

\begin{figure*}
\centering
\includegraphics[width=\textwidth]{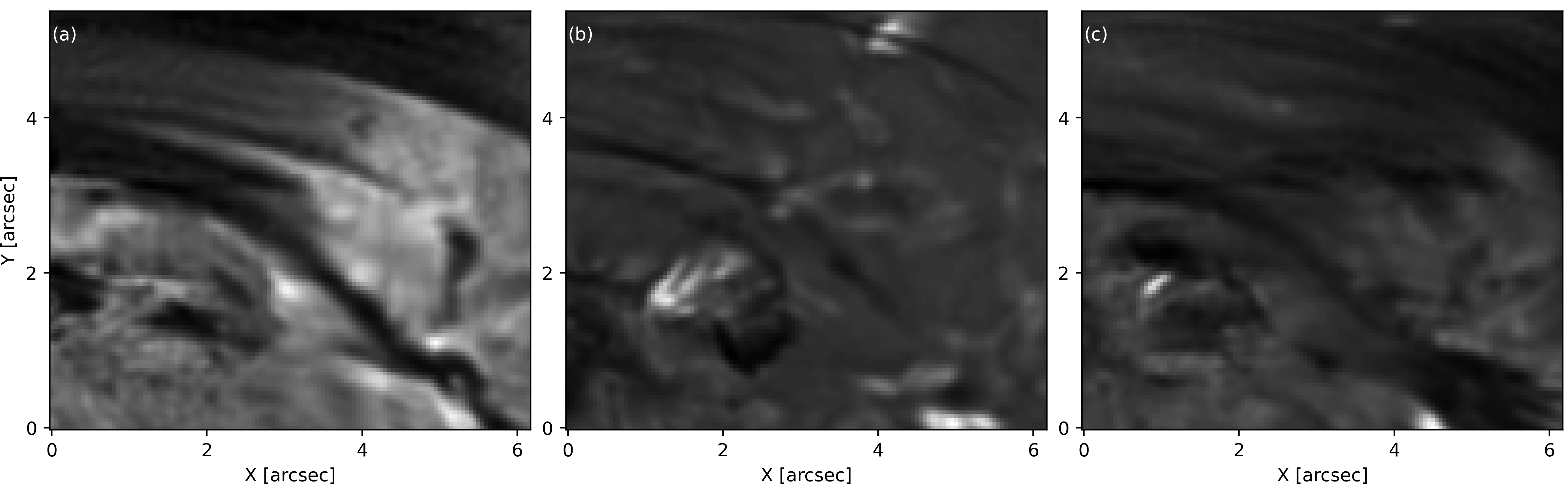}
\caption{ The first frame of the sequence of images taken by the MiHI obtained at the H$\alpha$ core position and integrated by $\pm$ 0.1 $\mathrm{\AA}$ (panel a). Panels (b) and (c) show the same frame taken at blue ($ \mathrm{H\alpha}-45\: \mathrm{km/s}$) and red wings ($  \mathrm{H\alpha}+50 \:\mathrm{km/s}$) correspondingly.   }
\label{fig:movie}
\end{figure*}

This paper is devoted to a case study of transverse oscillations detected in a fibril using a prototype Microlensed Hyperspectral Imager (MiHI) \citep{vanNoort2022A}. 

The manuscript is organised as follows. Section \ref{sec:obs} is devoted to the description of the instrument used, the dataset used for the analysis, and the complementary images used for the solar context. Section \ref{sec:methods} details the data analysis methods used to detect transverse waves in intensity maps and the derivation of Doppler velocities using the bisector method. Finally, Section \ref{sec:discussion} presents the interpretation and discussion of the results.

\section{Observations} \label{sec:obs}

The hyperspectral cube sequence used in the analysis was obtained with the MiHI on 24 August 2018, beginning at 08:03:44 UT, with a temporal cadence of 1.33 s and a duration of 10 min 6 s. 

The MiHI is an integral field spectrograph based on a double-sided microlens array (MLA), installed on an ordinary spectrograph instead of a slit. Following this principle, the prototype was executed as an extension of the TRI-Port Polarimetric Echelle-Littrow (TRIPPEL) spectrograph at the Swedish Solar Telescope \citep[SST;][]{Scharmer2003}. An integral field unit (IFU) can provide both spectral and 2D spatial information at high resolution simultaneously, which allows for a very high temporal cadence. In addition, compared to spectrographs and filtergraphs, IFUs can capture photons that are typically lost during spatial or spectral scanning, resulting in an increased photon efficiency. However, a disadvantage of IFUs is the challenge of achieving a sufficiently large field of view \citep{vanNoort2022C}. The use of IFUs in solar imaging has only recently been implemented (see the comprehensive review by \cite{Iglesias2019}), with several integral field solutions available. One such solution makes use of an MLA, as implemented by the MiHI. 
 
The adaptive optics (AO)-corrected science focal plane of the telescope is sampled by the MiHI's MLA, after which each sample is reduced in size to create a two-dimensional array of point-like sources. This array is then passed through the TRIPPEL spectrograph which disperses each point to generate a spectrum. The reduction and restoration techniques used to reduce the data are detailed in \cite{vanNoort2022C} and references therein. Due to difficulties in fitting an instrument model to the data \citep{vanNoort2022B}, an interpolation scheme was used to extract the data. This simple but approximate method leaves some spectral cross-talk in the data, which we corrected using an ad hoc correction scheme. Image restoration of the hyperspectral cube was performed according to \cite{vanNoort2017}, using atmospheric point spread functions (PSFs) recovered from a high-speed context camera. This approach resulted in high image stability, which reduced the need to compensate for jitter caused by atmospheric seeing or telescope rotation.

The wavelength range of the images is centred around the H$\alpha$ line, spanning 6560.56–6565.05 $\mathrm{\AA}$ with a spectral resolution of 10 $\mathrm{m\AA}$ px$^{-1}$. The pixel size is 0.065"px$^{-1}$. This particular dataset was previously used to investigate Ellerman bombs by \cite{RouppevanderVoort2023}. 

Fig. \ref{fig:movie} shows the first frame of the sequence. Panel (a) shows the MiHI image in the line core region, while panels (b) and (c) show images in the blue and red wings, respectively. In panel (a), the system of fibrils appears as dark, elongated threads crossing the whole field of view.  The images in both wings reveal the presence of Ellerman bombs that appear as compact, bright structures, alongside photospheric features such as pores.

\begin{figure*}
 \centering
\begin{tabular}{cc}
\includegraphics[trim={1cm 2cm 1cm 2cm},clip,width=.45\linewidth]{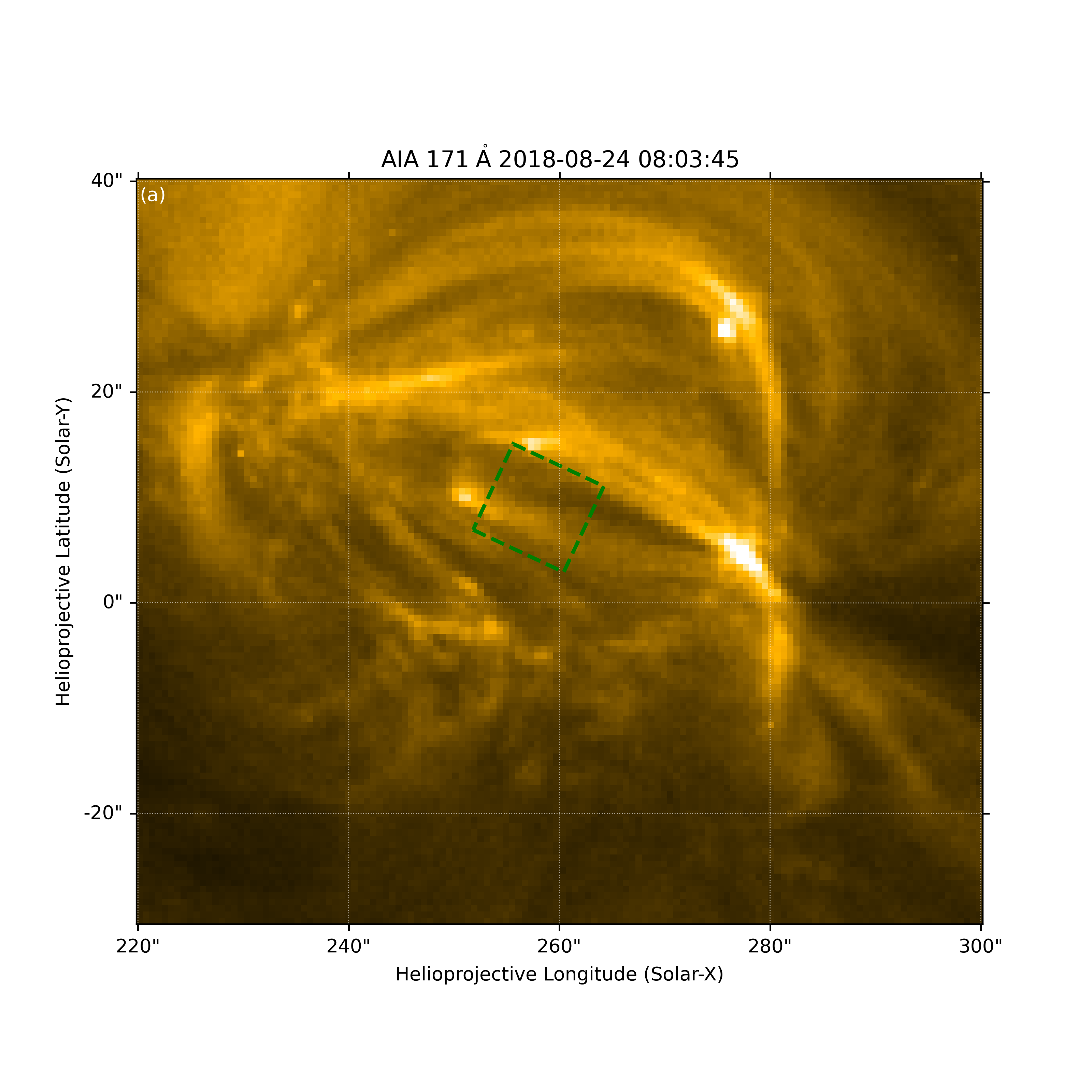} & \includegraphics[trim={1cm 2cm 1cm 2cm},clip,width=.45\linewidth]{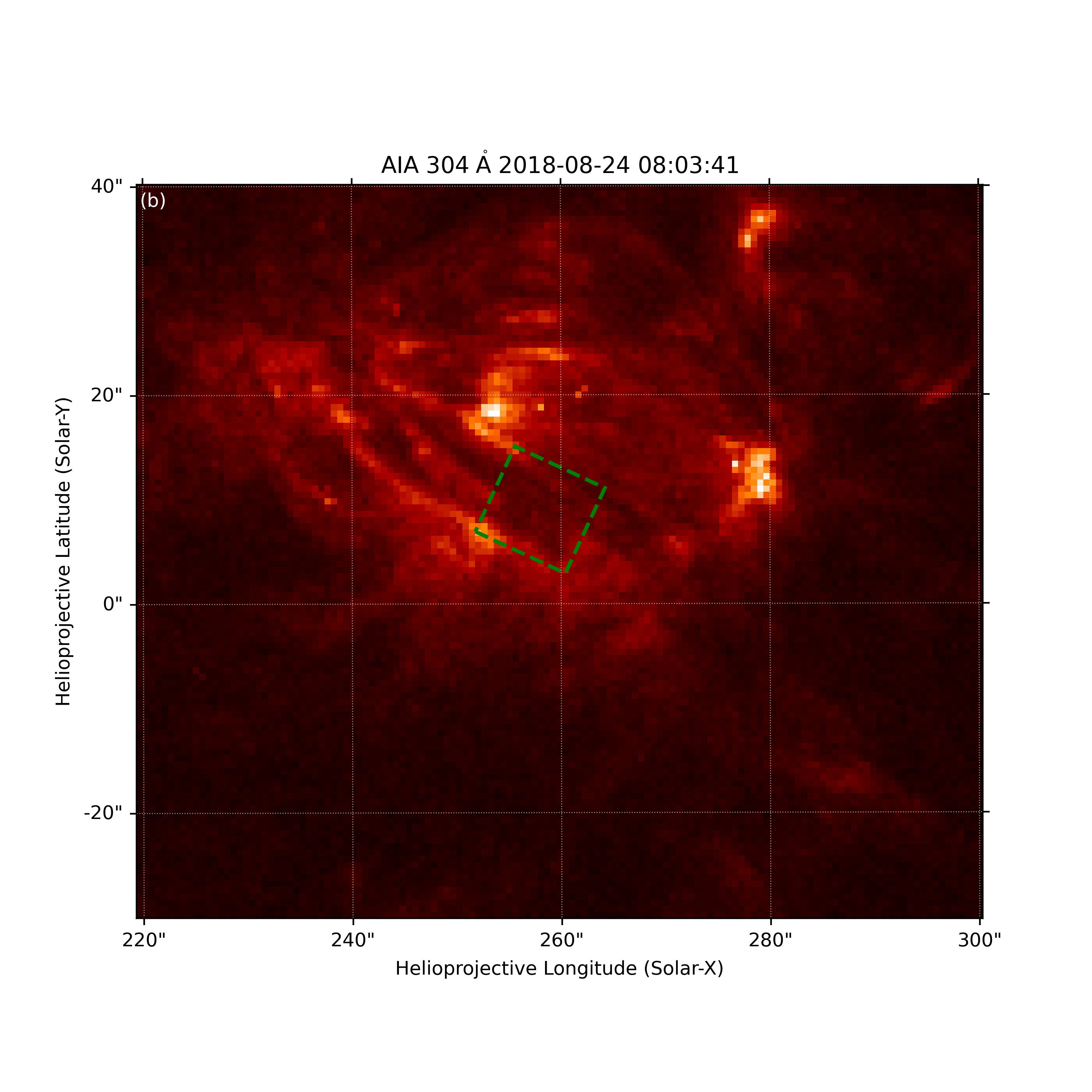} \\\includegraphics[trim={1cm 2cm 1cm 2cm},clip,width=.45\linewidth]{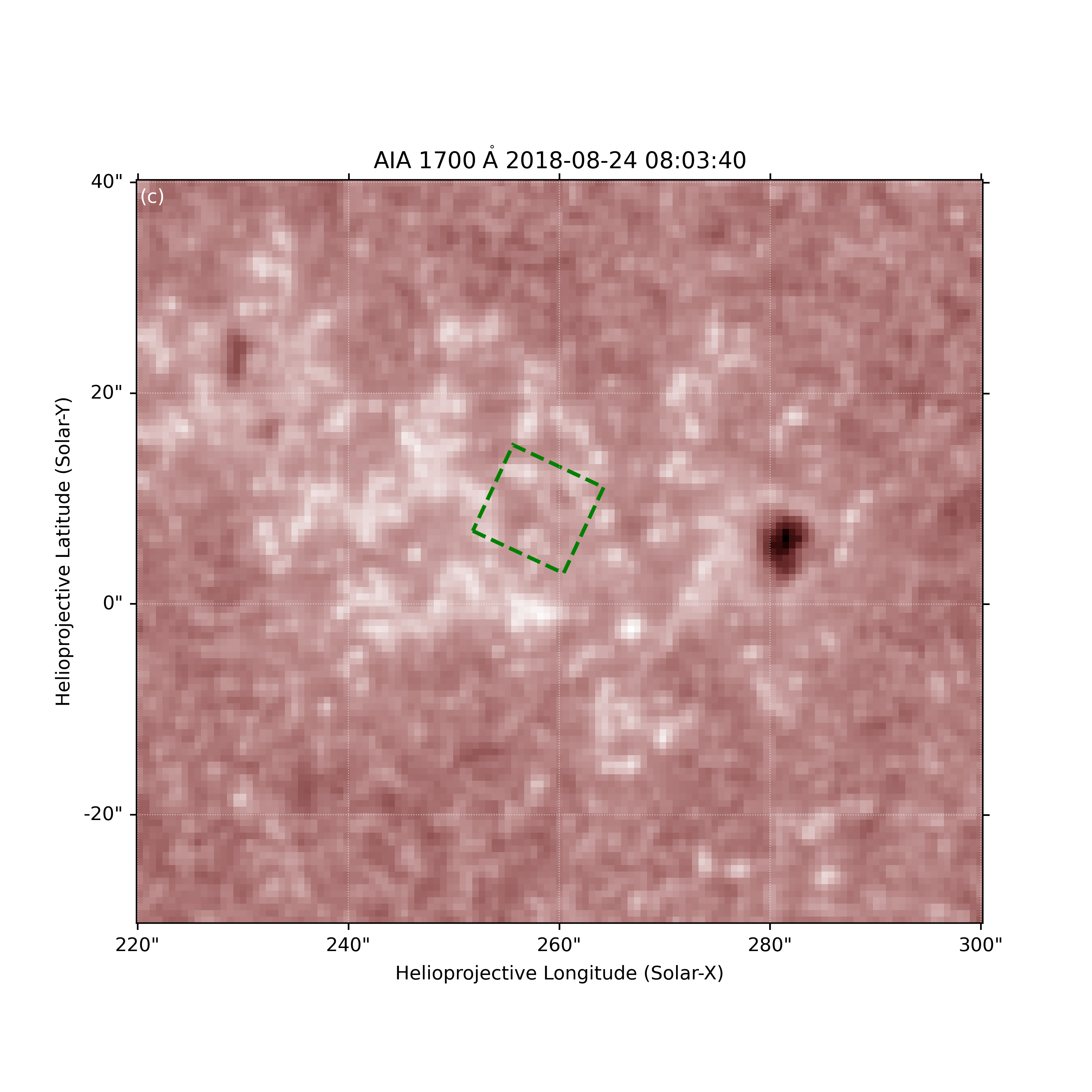} & \includegraphics[trim={1cm 2cm 1cm 2cm},clip,width=.45\linewidth]{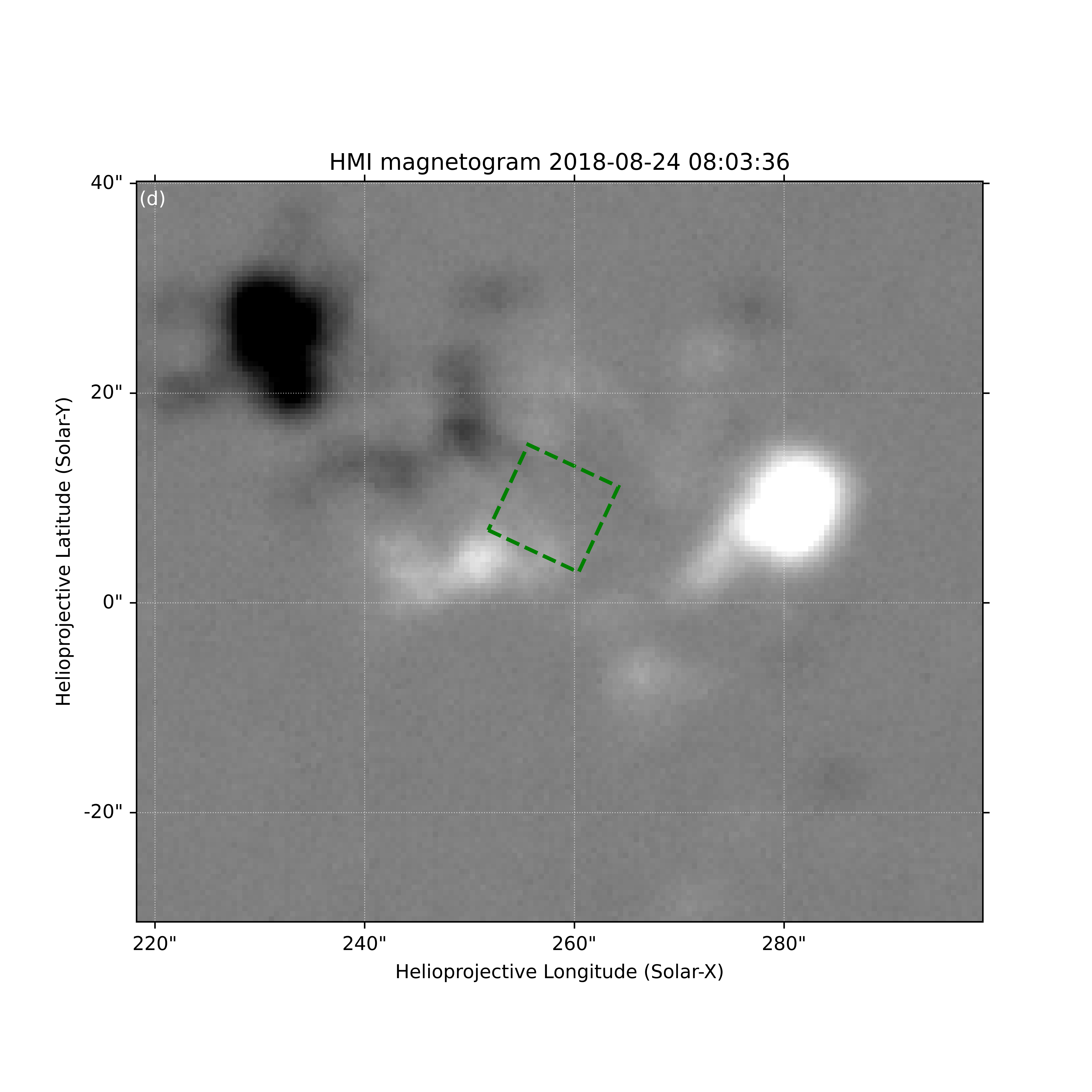}
\end{tabular}
\caption{AR\,12720 as observed by  the Solar Dynamics Observatory (SDO). Panels (a) to (c) show images in the 171 $\mathrm{\AA}$, 304 $\mathrm{\AA}$, and 1700 $\mathrm{\AA}$ channels. Panel (d) shows the LoS magnetic field in the photosphere obtained with the HMI. The green rectangle shows the field of vision (FoV) of MiHI. The resolution of the HMI magnetogram was reduced to match the Atmospheric Imaging Assembly (AIA) plate scale.  
}
\label{fig:aia}
\end{figure*}

The pointing centre of the observations was located at $\mathrm(x,y) = (255",5")$ in helioprojective coordinates, corresponding to the centre of the active region NOAA AR\,12720. Fig. \ref{fig:aia} shows images from the Solar Dynamics Observatory \citep[SDO;][]{Pesnell2012} taken using three channels of the Atmospheric Imaging Assembly \citep[AIA;][]{Lemen2012} and the Helioseismic and Magnetic Imager \citep[HMI;][]{Schou2012}. The images are the closest ones available to the beginning of the MiHI sequence. Panel (a) shows the AR in a coronal 171 $\mathrm{\AA}$ channel (log T = 5.8 K) with a bundle of coronal loops visible, while panel (b) shows the 304 $\mathrm{\AA}$ channel (log T = 4.7 K) that captures emissions from the upper chromosphere and transition region. Panel (c) presents the upper photosphere at the temperature log T = 3.7 K (1700 $\mathrm{\AA}$ channel) that shows several brightenings. The last panel (d) illustrates the magnetic field of the photosphere, with absolute values of the magnetic field saturated at $\pm$500 G. Magnetic concentrations corresponding to the footpoints of the coronal loops are clearly visible.

\section{Methods} \label{sec:methods}

\subsection{Data preparation}

Ground-based spectral observations are contaminated by telluric lines, which are produced by radiative transitions of the molecules that comprise the Earth's atmosphere. Telluric lines differ from solar lines because their widths are much narrower due to the low temperature of the Earth atmosphere \citep{vanNoort2022B}. In the observed spectral range, telluric lines are located approximately -78, +32, +57, and +64  $\mathrm{km \:s^{-1}}$ from the line core. 

\begin{figure}
\resizebox{8cm}{!}{\includegraphics{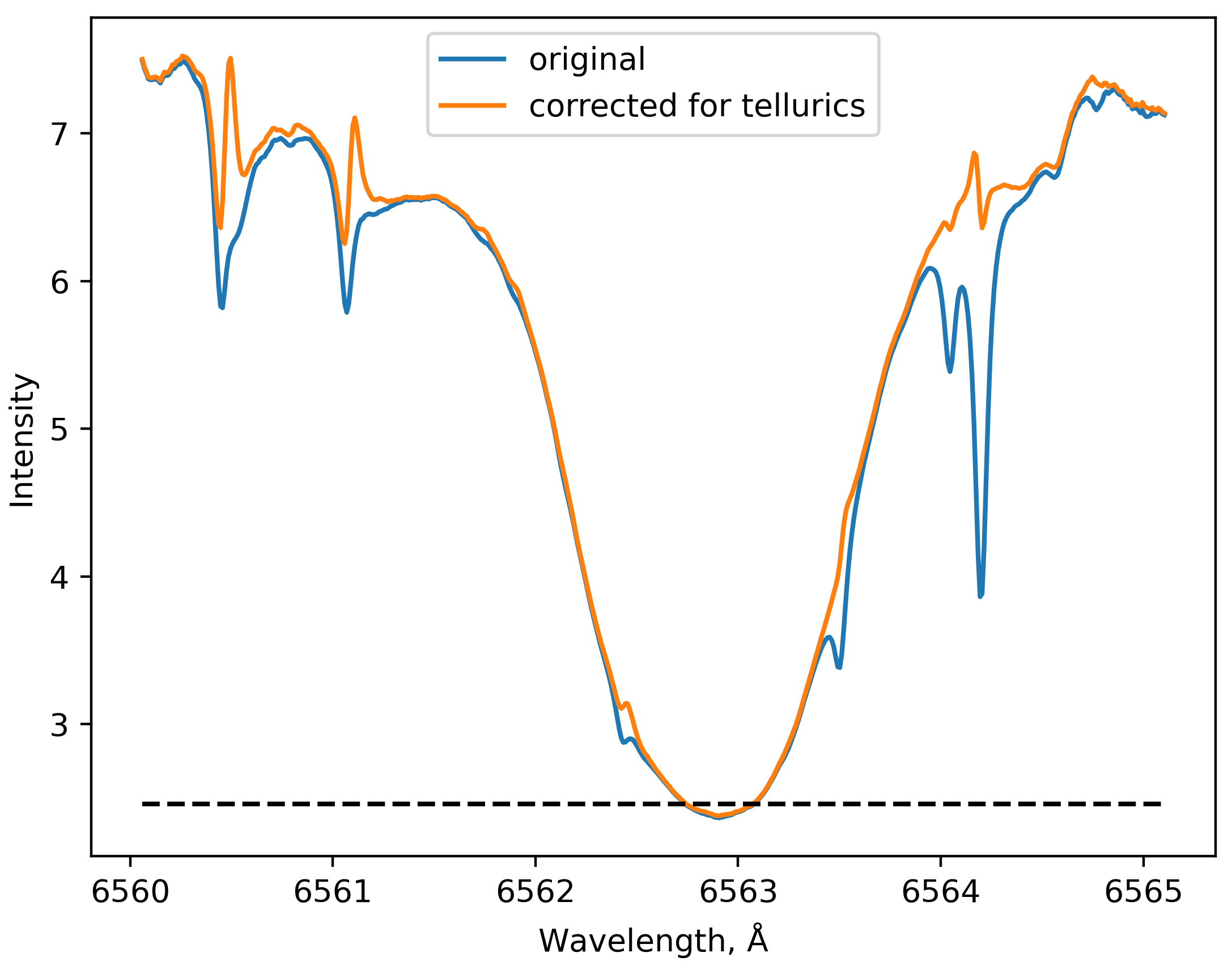}}
\caption{Mean profile constructed by averaging spectral lines for the whole FoV. The blue line shows the original profile. The orange line shows the mean profile corrected for tellurics. The black dashed line shows the level of the intensity used to construct Doppler maps. 
\label{fig:mean_profile}}
\end{figure}

To mitigate the effects of telluric contamination, the spectra were divided by the Kurucz telluric spectrum \citep{Kurucz2011} for the H$\alpha$ region, with a power of $\alpha$ ($\mathrm{spectrum} = \mathrm{spectrum}/\mathrm{tellurics}^\alpha$) where $\alpha$ varies from 3 to 5. The results are shown in Fig. \ref{fig:mean_profile}. The strength of the telluric lines is dependent on the elevation of the observatory and, in the case of water, on the weather.

\begin{figure}[!ht]
\resizebox{9cm}{!}{\includegraphics{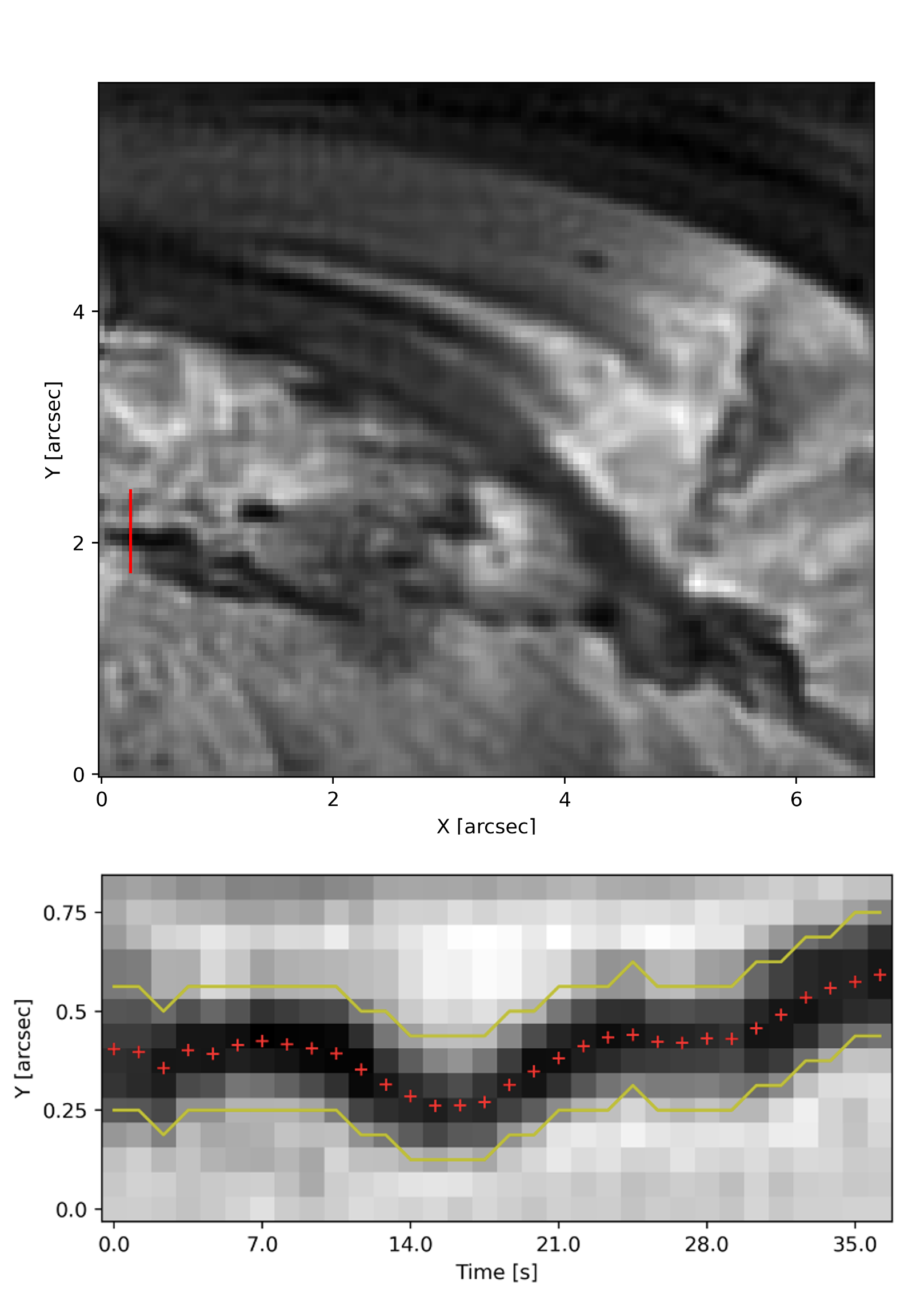}}
\caption{Intensity maps of the detected transverse oscillations. The upper panel shows the H$\alpha$ line core image, where the location of the pseudo-slit is denoted by the red line. The lower panel shows the time-distance map constructed using the red pseudo-slit. Red crosses show the centroid positions obtained through Gaussian fitting, while yellow curves show the fibril contour with a width of 5 pixels. 
\label{fig:td}}
\end{figure}

\subsection{Transverse waves in intensity signal}

We present a case study of one detected transverse oscillation event in a fibril observed in the core of the H$\alpha$ line. This particular region and fibril were chosen because they show a clear oscillatory signal in the motion of the fibril without overlapping with the motion of the other fibrils in the FoV. Transverse motions can be observed in the larger fibrils that cross the whole FoV, as shown in panel (a) of Fig. \ref{fig:movie}. However, their motion is highly overlapped, making consecutive fitting a very challenging task. The upper panel of Fig. \ref{fig:td} shows the pseudo-slit that was used to detect the oscillation signal. The lower panel displays the resulting time-distance map, where red dots indicate the position of the oscillating feature, identified by obtaining the centre position of a Gaussian fit to the intensity for each time frame. The yellow curves show the boundaries of the oscillating feature obtained by surrounding the obtained centre position with a 5-pixel width contour. We detected almost three oscillation cycles with a period $\mathrm{T}$ of 15 s and an amplitude $\mathrm{A}$ of 42 km, resulting in a velocity amplitude of $\mathrm{V_{\mathrm{Amp}}} = \frac {2 \pi A}{T} = 17.6$ $\mathrm{km \:s^{-1}}$. These characteristics were obtained by fitting the obtained centroid position with a sinusoidal function combined with an imposed background cubic trend. The detrending method was chosen empirically based on the dependence of amplitude variations on time. Experiments with different trends showed that the resulting fitting parameters are consistent with the quadratic fit. Although the linear fit yields a similar frequency parameter, it underestimates the amplitudes.

\begin{figure*}[!ht] 
\centering
\includegraphics[width = \textwidth]{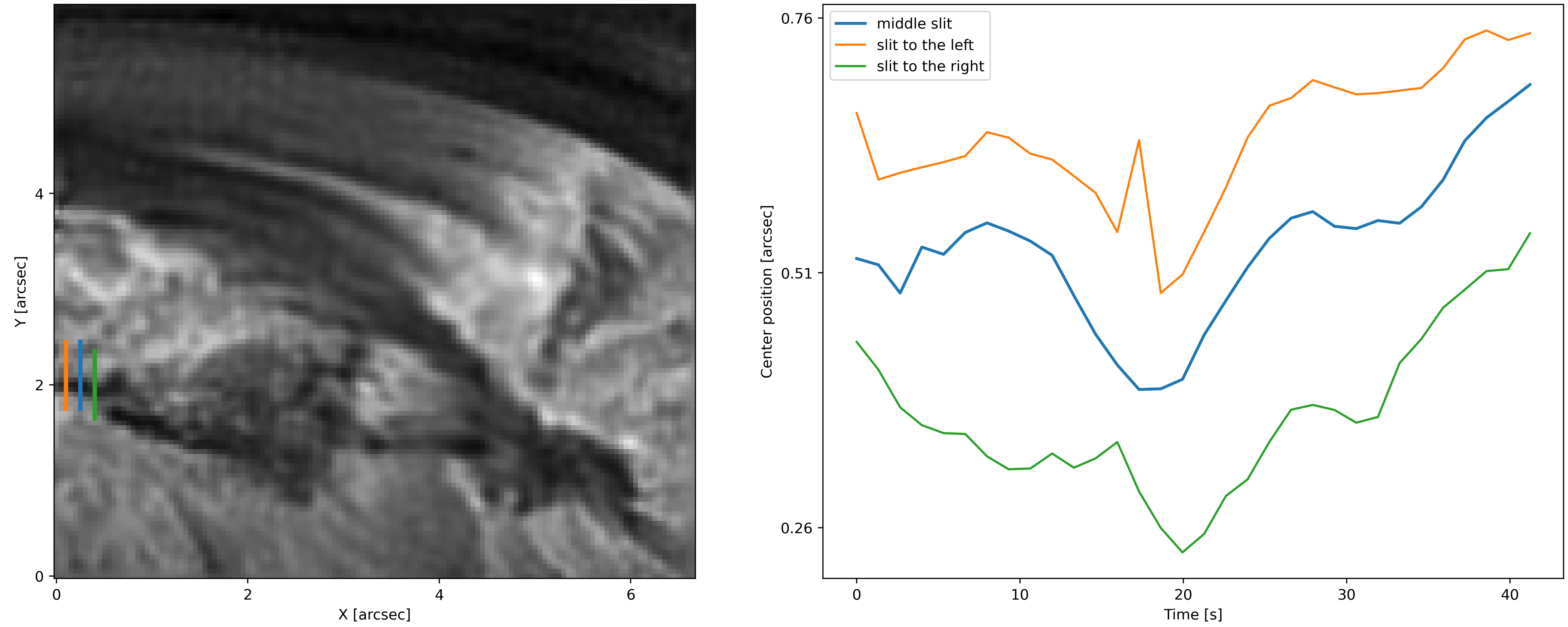}
\caption{Investigation of the propagation of the wave. The left panel shows the H$\alpha$ line core image, with the location of three pseudo-slits across the fibril. The right panel shows the detected centre positions of the fibril, with colours corresponding to the pseudo-slit used to construct the time-distance map.}
\label{fig:propagation}
\end{figure*}

Similarly to \cite{Shetye2021}, we investigated the phase relationship at different locations between the two ends and the midpoint of the fibril. The left panel of Fig. \ref{fig:propagation} illustrates the location of the three pseudo-slits at the positions mentioned. The panel on the right shows the detected oscillating fibril position, where the colour of the curve corresponds to the colour of the pseudo-slit used to construct a time-distance map and detect the centre position of the fibril. The absolute position on the right panel does not carry physical significance; only the relative positions can be considered. This is because the plot was designed to ensure that all curves are separated and clearly visible, allowing for the comparison of phase relationships.

The cross-correlation analysis was performed via cross-correlation of two continuous wavelet transforms using a Morlet mother wavelet. The cross-wavelet transform plots reveal (see Fig. \ref{fig:appendix4} in Appendix \ref{appendix2}) in-phase behaviours between the displacement of the middle slit versus the slits to its left and right. Additionally,  an apparent propagation was visible for the minima located around 20 s. Panel (b) of  Fig. \ref{fig:propagation} visually illustrates that there is no phase shift and no significant difference in amplitude. The detected wave is not necessarily a standing wave. It could be a propagating wave if the propagation speed is relatively high. The distance between the pseudo-slits is 3 pixels (150 km), which means that for the cadence of 1.33 s, we are only able to detect the propagating velocities slower than approximately 110 $\mathrm{km \:s^{-1}}$.

\subsection{Bisector method}

To calculate Doppler velocities, we applied one of the most widely used methods, the bisector method \citep{Kulander1966}. The bisector method relies on full line profiles and uses their asymmetries, which are normally associated with the velocity gradients along the LoS. The line bisector is constructed from points that, at a given intensity level, are in the middle of the profile. The bisector shift is then calculated as the distance from the static line centre to the line bisector, which gives then the velocity at different atmospheric layers ranging from the line core to the wings.

The bisector method has been proven to be reliable for some lines, such as the upper photospheric \ion{Si}{I} 10827 $\mathrm{\AA}$ line. This reliability was demonstrated using Bifrost simulations by comparing the inferred velocities of bisectors taken at different line profile intensities and simulation velocities at given optical depth\citep{Gonzalez2020}. It has also been extensively used for the chromospheric \ion{Ca}{II} 854.2 nm line \citep[e.g.][]{Leenaarts2014,Pietarila2013} and H$\alpha$ lines \citep{Chae2013, Maurya2013}. However, one limitation of this method is the difficulty in assigning velocities to a specific layer of the atmosphere. The derived Doppler velocities contain information from a large portion of the atmosphere, making it challenging to pinpoint the exact layer that is being measured.

Additional difficulties in deciphering bisector results for the H$\alpha$ line arise from the lack of understanding of its formation, due to the complicated physics of the chromosphere.  \cite{Leenaarts2012} used radiation MHD simulations to show that H$\alpha$ intensity in the core region of the line is correlated with the average formation height:  the higher the height, the lower the line core intensity.

When applying the bisector method, the profile is typically fitted using variations of Gaussian \citep{Gilchrist-Millar2021}, Lorentzian or Voigt functions \citep{voigtexample_2019}. However, these functions often fail to provide the best fit, as real profiles can deviate significantly from these models. This can make determining LoS velocities challenging. For instance, regions with a strong magnetic field often have broadened lines due to Zeeman splitting \citep{Solanki1993} or line weakening \citep{Rubio2019}. The mentioned functions consequently do not accurately describe the profiles of these lines.

This is not the only challenge in such regions, as quiet areas also often exhibit broad line profiles with a flattened plateau around the line core \citep{Leenaarts2012}. One possible approach to address this issue is to use the supersech function, as described in Appendix \ref{appendix1}.

We examined profiles from pixels outside the analysed fibril location and its close surroundings, which revealed a clear signature of multiple components. These features are detectable only due to the high spectral resolution of the MiHI instrument and can be missed with a lower spectral resolution (see, for example, Fig. \ref{fig:appendix3} in the Appendix \ref{appendix1}).

Various methods can be used to derive Doppler velocities, some of which have been tested, as described in the Appendix. In the current section, we used the bisector method on the original profiles fitted with an asymmetric Gaussian function, focusing primarily on phase investigation for polarisation studies.

\begin{figure}
\resizebox{9cm}{!}{\includegraphics{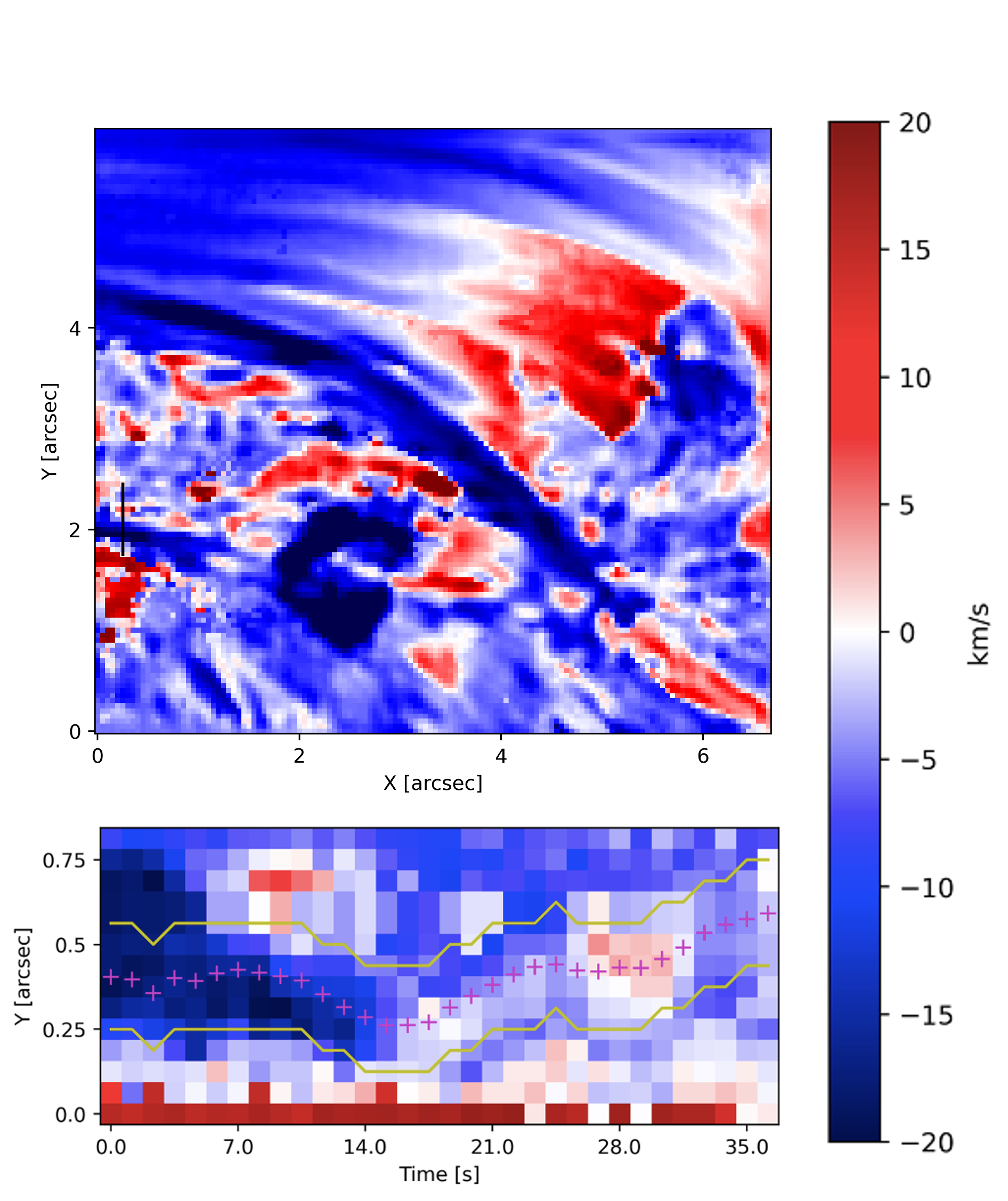}}
\caption{Doppler maps of the detected transverse oscillations. The upper panel shows the Doppler velocity map obtained using the bisector method. The black line shows the location of the pseudo-slit (same as in Fig. \ref{fig:td}) that is used to construct the Doppler time-distance map shown in the lower panel. Magenta crosses show the fibril position as obtained from the intensity map and yellow curves outline the fibril. 
\label{fig:td_doppler}}
\end{figure}

The Doppler map derived using the bisector method is shown in the upper panel of Fig. \ref{fig:td_doppler}. This map illustrates the distribution of Doppler velocities calculated for the first image of the sequence, taken at an intensity level of 2.46 which belongs to the line core (denoted by the black dashed line in Fig. \ref{fig:mean_profile}). However, due to the complex behaviour of H$\alpha$ formation, it is difficult to map it to a certain height in the chromosphere. White pixels represent cases where the algorithm failed, either due to profile fitting issues (e.g. the profiles were fitted using a Gaussian function) or the absence of a signal at the selected intensity value.

The lower panel of Fig. \ref{fig:td_doppler} shows the Doppler velocities time-distance map, constructed similarly to the intensity map, with the same centre positions and fibril contours overplotted. A flow is present in the detected signal, which will be subtracted when investigating the polarisation of the detected oscillation. Fourier analysis of the centre position reveals a signal period of 14.2 seconds.

\begin{figure*}[h!]
    \centering
    {\includegraphics [width = \textwidth]{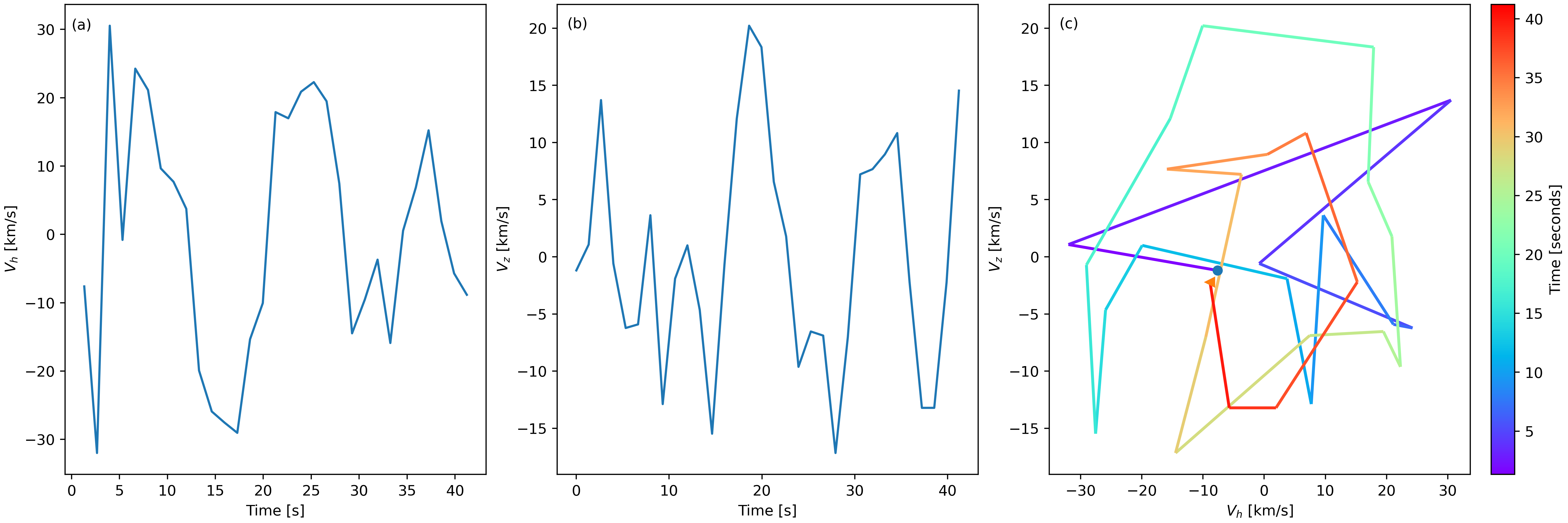}}
     \caption{Original signal of the velocity oscillations with the detrending applied. Panel (a) and (b) show PoS and LoS velocities, respectively. Panel (c) shows the hodogram where the colour represents the time. The starting point is denoted by the blue dot and the ending point is denoted by the red triangle.}
      \label{fig:hodogram_original}
\end{figure*}

\begin{figure*}[!ht]
\centering
\includegraphics[width = \textwidth]{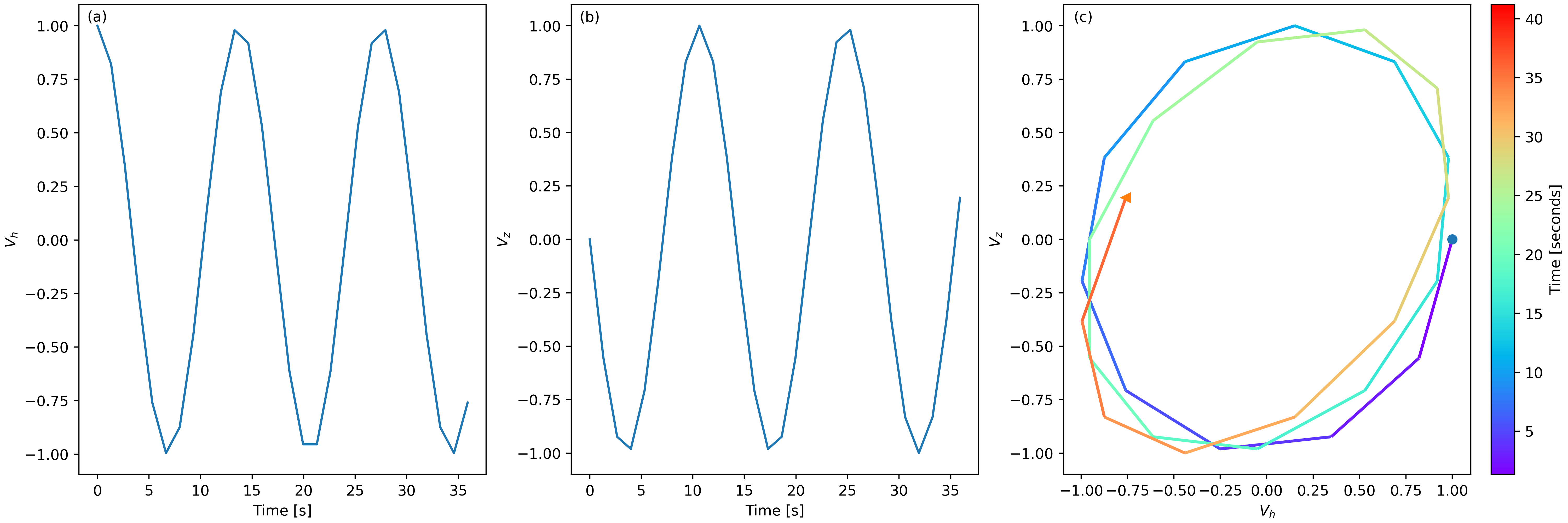}
\caption{Filtered signal of the velocity oscillations. Panel (a) and (b) show the normalized PoS and LoS velocities to their maximum values, respectively. Panel (c) shows the hodogram where the colour represents the time. The starting point is denoted by the blue dot and the ending point is denoted by the red triangle.} 
\label{fig:hodogram_filtered}
\end{figure*}

\subsection{Polarization of the transverse wave}

To illustrate the polarisation of transverse waves, we constructed a phase portrait, or hodogram, following the approach of \cite{Zhong2023,Bate2024}, which is a particularly useful tool for demonstration purposes.  Panel (a) of Fig. \ref{fig:hodogram_original} shows the plane of sky (PoS) velocities ($V_h$), calculated as the derivative of the centroid positions shown in Fig. \ref{fig:td}, with the long-term trend removed. Similarly to the amplitudes in the transverse direction, we removed the long-term trend, which corresponds to the background flow,  from the Doppler velocities ($V_z$) using a cubic fit. The resulting data are shown in panel (b) of Fig. \ref{fig:hodogram_original}. 

Panel (c) shows the hodogram where $V_h$ is plotted against $V_z$ as they progress with time. The colour of the curve represents how much time in seconds has elapsed since the detection of the oscillation signal in the first frame. By plotting two velocity components in perpendicular directions, the shape of the figure should provide insights into the polarisation of the wave. Although the hodogram is challenging to interpret, three clockwise rotations can be observed from the starting point (dot) to the end point (triangle). To facilitate interpretation, we filtered all frequencies except the main oscillation frequency of $\sim$15 s, which has the highest power. Apart from that, we also remove any amplitude modulation. Panels (a) and (b) of Fig. \ref{fig:hodogram_filtered} show the results of all filtering procedures applied to velocities in the PoS and LoS directions.

Panel (c) of Fig. \ref{fig:hodogram_filtered} shows the hodogram for the filtered velocities, where the shape is clearly prominent and exhibits an elliptical form. This indicates that the polarisation of the detected transverse wave is also elliptical. This result is consistent with observations reported by \cite{Shetye2021} and \cite{Bate2024}.

\section{Discussion and conclusion} \label{sec:discussion}

We conducted a case study of an oscillating fibril using the MiHI instrument, revealing the following characteristics of the wave: a period of approximately 15 s, a displacement amplitude of 42 km, and a velocity amplitude of 17.6 $\mathrm{km \:s^{-1}}$. These parameters are consistent with previous observations, as reported by \cite{Kianfar2022, Jafarzadeh2017}. By combining Doppler velocities deduced via the bisector method and PoS velocities, we were able to characterise the polarisation of the detected wave as elliptical. This elliptical polarisation is in agreement with the statistical studies performed by \cite{Shetye2021,Bate2024}.  

The estimated energy content of the oscillations remains uncertain because of the challenges of attributing the observed phenomenon to a specific wave mode. However, Fig. \ref{fig:td_doppler} (b) shows no evidence of torsional motions, and therefore the kink mode is the most likely interpretation. For this scenario, we estimate the energy flux using the following formula \citep{Bate2024}:

\begin{equation}
	F \approx \frac{1}{2}f\rho_iV_{\mathrm{Amp}}^2V_{\mathrm{ph}}.
\end{equation}

This formula is a simplified version of that derived for kink waves by \cite{Doorsselaere2014}
We adopt an internal fibril density of $\rho_i$ = $10^{-9}-10^{-10}$ $\mathrm{kg\: m^{-3}}$,  taken from simulations of a model chromosphere by \cite{Leenaarts2012}, and a filling factor, $f$, of 5$\%$ \citep{Bate2024}. 
The absence of phase lag precludes the use of established phase lag analysis methods to estimate the phase speed $V_{\mathrm{ph}}$. We therefore adopt values from the literature that span a range of $50 - 300$ $\mathrm{km \:s^{-1}}$ \citep{Kianfar2022}. This results in a range of values for the energy flux of 390 - 2300 $\mathrm{W\:m^{-2}}$. Compared with the total chromospheric losses of $2\cdot10^4$ $\mathrm{W\:m^{-2}}$ for active regions \citep{Withbroe1977}, it is evident that even at the upper limit, this energy flux is not sufficient to compensate for the losses.

The power law of waves in the chromosphere, constructed by \cite{Lim2023} for transverse waves in the corona, requires further investigation. Analysing a single event is insufficient to construct a robust power law and identify the frequency ranges with a higher energy budget that could compensate for the heating.  A statistical analysis of multiple events would be beneficial for understanding the frequency range with a higher energy budget that could potentially compensate for the chromospheric losses. This statistical approach could serve as the next step in investigating the energy dynamics of transverse waves in the chromosphere.

\begin{acknowledgements}
The authors thank the referee for their valuable comments. 
TVD was supported by a Senior Research Project (G088021N) of the FWO Vlaanderen. Furthermore, TVD received financial support from the Flemish Government under the long-term structural Methusalem funding program, project SOUL: Stellar evolution in full glory, grant METH/24/012 at KU Leuven. The research that led to these results was subsidised by the Belgian Federal Science Policy Office through the contract B2/223/P1/CLOSE-UP. It is also part of the DynaSun project and has thus received funding under the Horizon Europe programme of the European Union under grant agreement (no. 101131534). Views and opinions expressed are however those of the author(s) only and do not necessarily reflect those of the European Union and therefore the European Union cannot be held responsible for them. EP has benefited from the funding of the FWO Vlaanderen through a Senior Research Project (G088021N).
\end{acknowledgements}

\bibliography{References}{}
\bibliographystyle{aa}

\newpage
\appendix 

\section{Deriving Doppler velocities } \label{appendix1}

We have explored different methods to derive Doppler velocities, as the bisector method has its limitations due to the underlying assumption that the moving structure generates its own line profile, which is shifted relative to the background \citep{Alissandrakis1990}. It has been shown below that this assumption does not always hold.

   \begin{figure}[h!]
   \centering
   \includegraphics[width=\hsize]{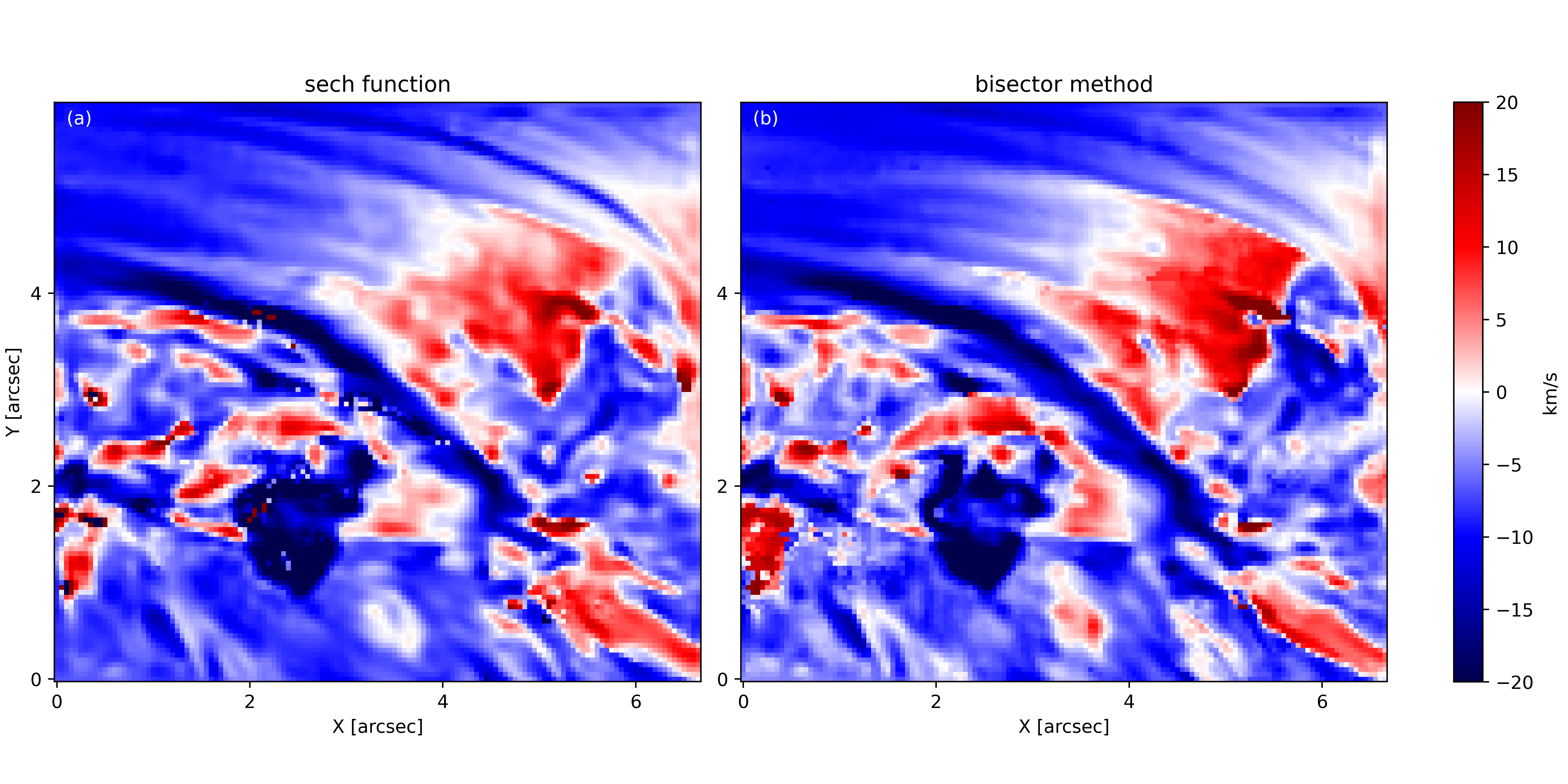}
      \caption{Doppler velocity maps. Panel (a) shows the velocities using the super-sech function. Panel (b) shows the velocities derived with the bisector method.}
         \label{fig:appendix1}
   \end{figure}

      \begin{figure}[h!]
   \centering
   \includegraphics[width=\hsize]{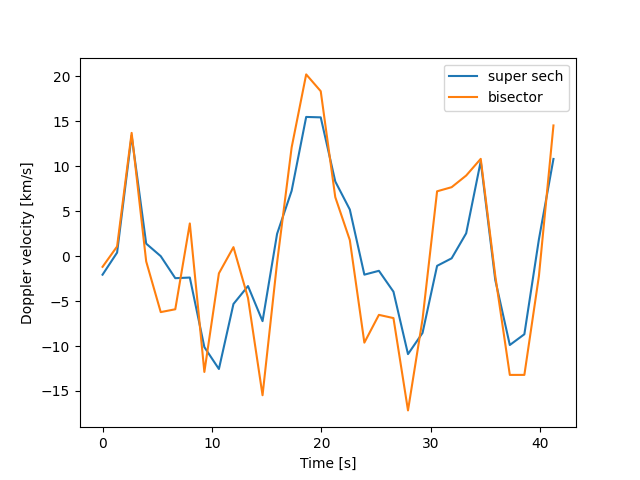}
      \caption{Comparison of the averaged Doppler velocities for the oscillating fibril. The orange curve shows the velocities derived using bisector method. The blue curve shows the Doppler velocities obtained using the super-sech function. }
         \label{fig:appendix2}
   \end{figure}

\begin{figure}[h!]
   \centering
   \includegraphics[width=\hsize]{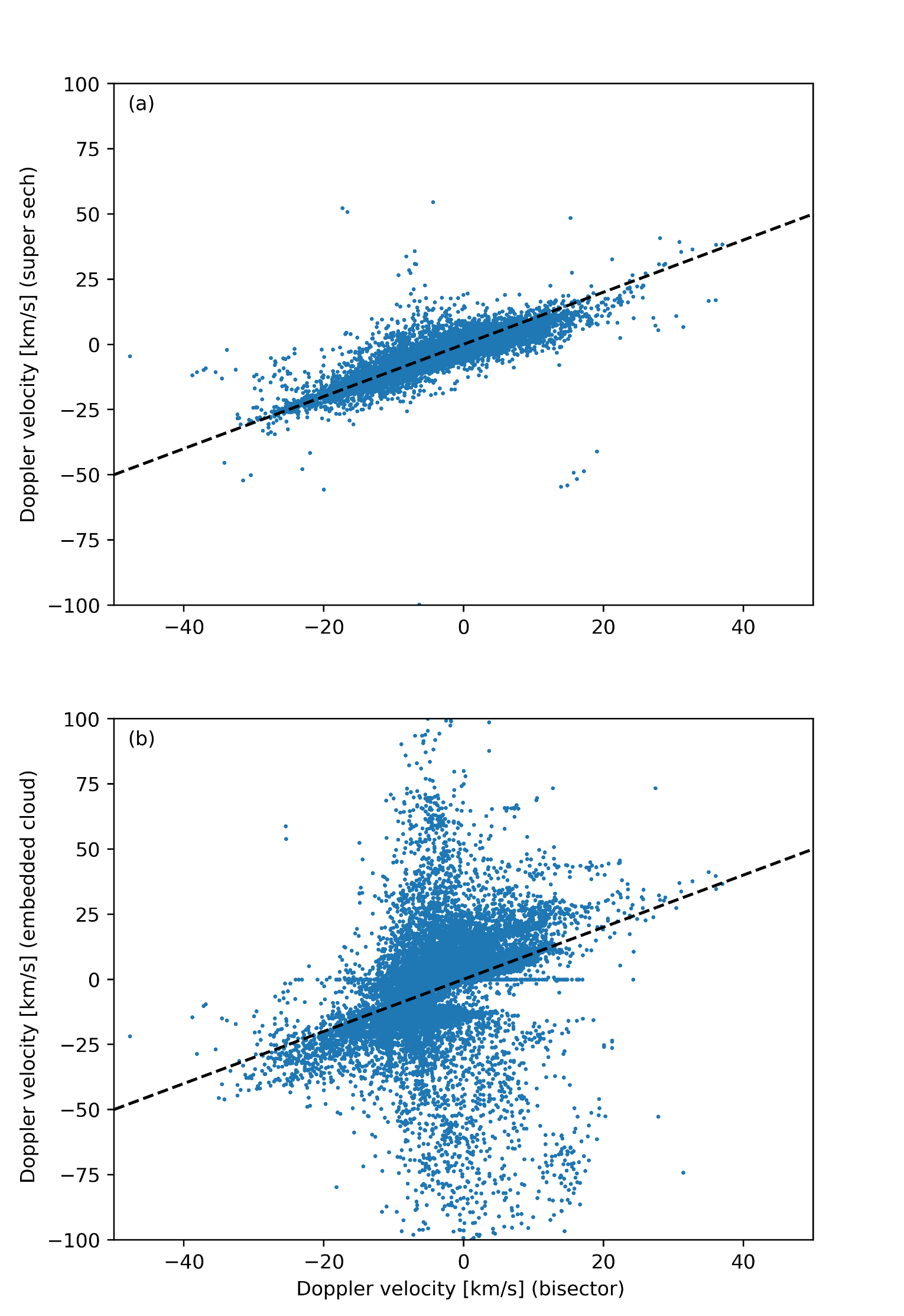}
      \caption{Correlation plots of Doppler velocities of the first frame of the image sequences obtained using different methods. Panel (a) shows Doppler velocities obtained using bisector method plotted against velocities obtained using super-sech fit. Panel (b) shows Doppler velocities obtained using bisector method plotted against velocities derived using embedded cloud model. In both panels the black dashed line shows $x = y$ line.  }
         \label{fig:correlation}
   \end{figure}

An additional method we tried was using the super-sech function, as in \cite{Bate2024}, which has been shown to provide a better fit than Gaussian or Voigt functions for profiles with a flat core. The resulting Doppler map derived from the first frame of the sequence is shown in Fig. \ref{fig:appendix1} (b) and appears smoother compared to the original method (panel (a)), while overall resembling the originally derived Doppler map with some minor differences. However, the values are not significantly different when examining the averaged Doppler velocities at the location of the analyzed oscillating fibril (shown in Fig. \ref{fig:appendix2}). Therefore, we used the values derived using the bisector method in the analysis. Panel (a) of Fig. \ref{fig:correlation} shows a scatter plot of Doppler velocities derived using the two methods and shows overall good correspondence of the velocity values.

   \begin{figure}[h!]
   \centering
   \includegraphics[width=6cm]{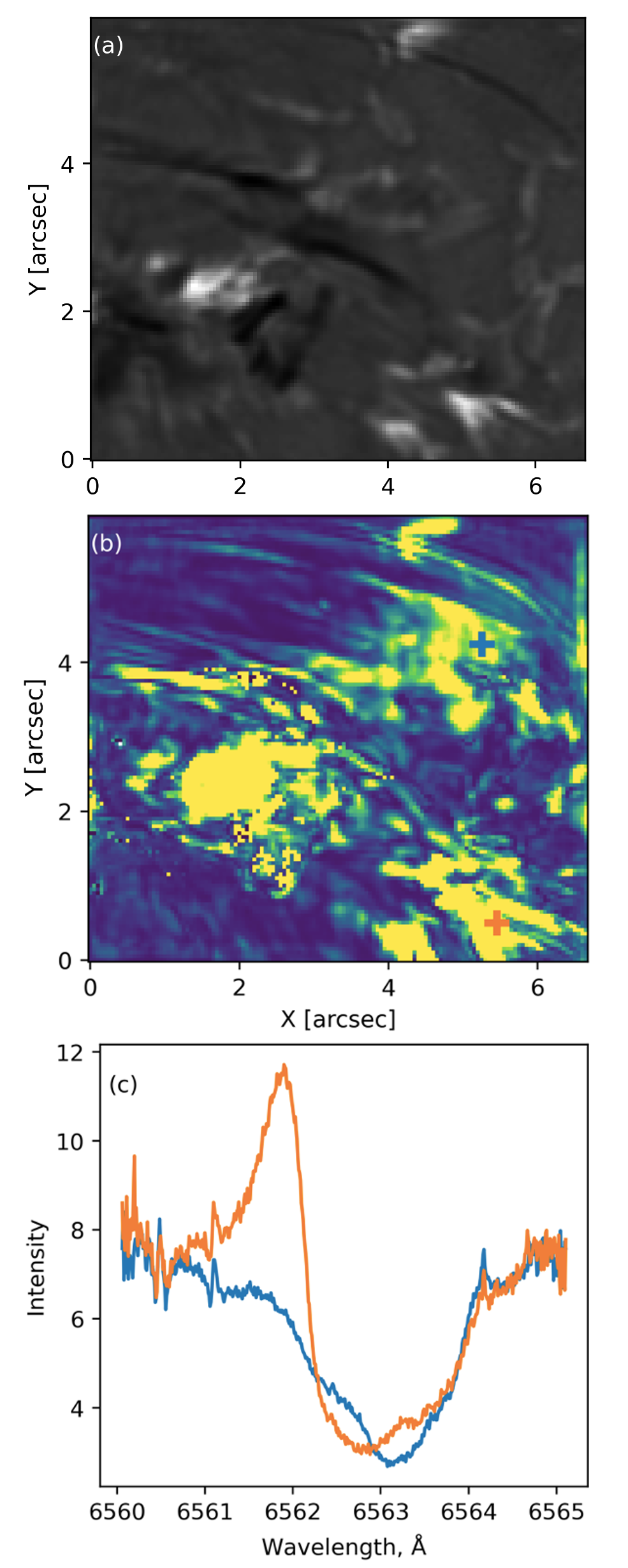}
      \caption{Panel (a) shows the intensity map taken at the blue wing position of H$\alpha$. Panel (b) shows the $\chi^2$-error map of the super-sech fit. Blue and orange crosses show the locations of pixels which spectra are shown on panel (c) where colour of the curve corresponds to the colour of the cross. }
         \label{fig:appendix3}
   \end{figure}

The $\chi^2$-error map of the fit, calculated excluding the continuum, is shown in Fig. \ref{fig:appendix3} (b). While the location of the fibril appears unaffected, and both methods are suitable for deriving velocity values, they are not the most reliable for the entire FoV. 

The reason for this, as seen in panel (c) of Fig. \ref{fig:appendix3}, is that the profiles deviate from the typical Gaussian or Lorentzian shape either by having an additional component present such as the one shown by the blue curve, or strong emission present as shown by the orange curve. Furthermore, it can be noted that the regions of higher error values correspond to the shapes seen in images taken at the blue wing position (Fig. \ref{fig:appendix3} (a)), where emission from the lower atmosphere is clearly visible. 

Therefore, we also tested methods that do not assume an initial profile shape, specifically `cloud models. In this study, we used the original Beckers cloud model \citep{Beckers1964} and the embedded cloud model \citep{Chae2014}, with modifications to account for a variable source function.

However, due to the shape of the profiles, the model often fails, being unable to fit the contrast profiles using the 4 free parameters in the original model and the 9 parameters (with 3 fixed) in the embedded model. As a result, for pixels located within the outlined contour, the fit fails in 40\% of the cases, particularly in regions with a clear emission component in the blue wing. In these cases, the resulting profile cannot be described by any of the contrast profiles provided by \cite{Grossmann-Doerth1971}. Panel (b) of Fig. \ref{fig:correlation} shows clear difference between the velocities derived using bisector method and embedded cloud model.

Therefore, when fitting the entire field of view, it may be best to consider more complex cloud models, such as those described by \cite{Chae2020}, or to explore recently developed machine learning techniques like those in \cite{MacBride2021}. The latter technique can separate multiple components in the spectra using neural networks, allowing for the derivation of Doppler velocities for the component of interest. 

\section{Cross-wavelet analysis}\label{appendix2}

\begin{figure}[h!]
   \centering
   \includegraphics[width=\hsize]{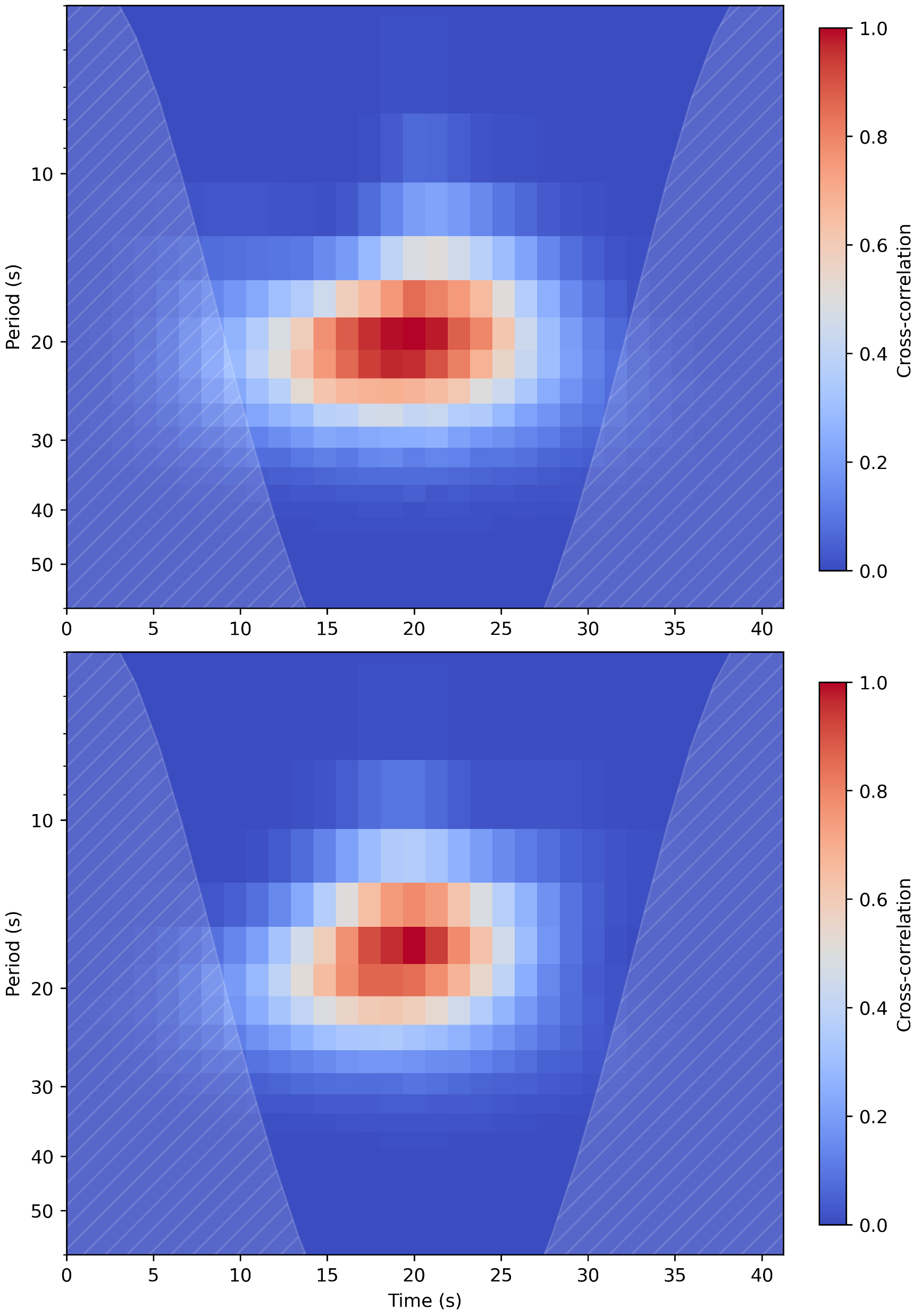}
      \caption{The cross-wavelet correlation between two transverse motions is shown. The upper panel displays the cross-correlation between the middle slit and the slit to its left, while the lower panel shows the cross-correlation between the middle slit and the slit to its right. Red indicates regions of maximum significance, while blue represents regions of minimum significance. }
         \label{fig:appendix4}
   \end{figure}

Fig. \ref{fig:appendix4} shows the cross-wavelet correlation between the oscillation signals in different slits. Both panels show in-phase behavior.

\end{document}